\documentclass[a4paper,10pt]{scrartcl}
\pdfoutput=1
\usepackage[latin1]{inputenc}
\usepackage[ngerman, english]{babel}
\usepackage{graphicx}
\usepackage{setspace}
\usepackage{lmodern}
\usepackage{multicol}
\usepackage{csquotes}
\usepackage{hyperref}
\usepackage[natbib]{biblatex}
\bibliography{SocialNetworks2}
\usepackage[a4paper,head=3cm,headsep=15mm,height=242mm,bottom=25mm,left=4cm,right=3cm]{geometry}
\hypersetup{pdfauthor={Jörg Pohle},pdftitle={Social Networks, Functional Differentiation of Society, and Data Protection},pdfkeywords={privacy, data protection, right of personality, functional differentiation, separation of duties, balance of power, social networks}}

\begin{document}
	
\selectlanguage{english}

\title{Social Networks, Functional Differentiation of Society, and Data Protection}
\author{Jörg Pohle\thanks{Humboldt University at Berlin, Institute of Computer Science, Computer Science in Education and Society, \texttt{pohle@informatik.hu-berlin.de}.}}

\pagestyle{plain}

\maketitle

\begin{abstract}
\noindent\textbf{Abstract.} Most scholars, politicians, and activists are following individualistic theories of privacy and data protection. In contrast, some of the pioneers of the data protection legislation in Germany like Adalbert Podlech, Paul J. Müller, and Ulrich Dammann used a systems theory approach. Following Niklas Luhmann, the aim of data protection is (1) maintaining the functional differentiation of society against the threats posed by the possibilities of modern information processing, and (2) countering undue information power by organized social players. It could be, therefore, no surprise that the first data protection law in the German state of Hesse contained rules to protect the individual as well as the balance of power between the legislative and the executive body of the state.

Social networks like Facebook or Google+ do not only endanger their users by exposing them to other users or the public. They constitute, first and foremost, a threat to society as a whole by collecting information about individuals, groups, and organizations from different social systems and combining them in a centralized data bank. They transgress the boundaries between social systems that act as a shield against total visibility and transparency of the individual and protect the freedom and the autonomy of the people. Without enforcing structural limitations on the organizational use of collected data by the social network itself or the company behind it, social networks pose the worst totalitarian peril for western societies since the fall of the Soviet Union.
\end{abstract}

\begin{abstract}
\end{abstract}

\section{Introduction}

General social networking services\footnote{Specialized networks like LinkedIn and XING will not be surveyed.} are used by more than a billion users worldwide. They produce a large volume of data about their users, either collected directly from the individuals themselves or derived from the social graph around each of them. The information contained in the data relates to all areas of life, all social systems, and the past, the present, and the future alike. Additionally, they often also relate to individuals unaffiliated with said social networks. Social networks are often harshly criticized from privacy activists, data protection commissioners, and politicians alike for exposing their users to the public, for their low privacy and security standards, and their incomprehensible privacy policies.

The aim of this paper is to show that the danger of social networks to individuals and society alike are better understood from a structuralist point of view than from the individualist one used by most scholars, politicians, and activists.

In Chapter 2, I begin with a brief survey of the history and the three main lines of the modern information privacy and data protection discourse. Even though the history of privacy regulation can be traced back to the Hippocratic oath and the secrecy of confession, modern theories of privacy appear in the second half of the nineteenth century. A hundred years later, the right to privacy and the general right of personality are accepted legal concepts in the US and in Germany, respectively. Against this background the advent of the computer and the electronic data processing ignites the public debate on information privacy. Most scholars follow one of two traditional privacy theories: (1) privacy as secrecy and confidentiality, or (2) privacy as right of personality and individual self-determination. In contrast, some of the pioneers of the data protection legislation in German use an approach based on concepts borrowed from Luhmann's sociological systems theory (3).

In Chapter 3, I give a short overview of the concept of functional differentiation used to characterize social systems in modern western societies. I show how society, social systems, and the individual benefit from functional differentiated social systems, and how modern information processing threatens these achievements. I then present the aim of data protection as envisioned by some of the pioneers of the German data protection legislation.

Finally, in Chapter 4, I examine the effects of general social networking services on the functional differentiation of society, the balance of power between social systems, and the individual's chance for self-determined role-playing. I then demonstrate how social networks and the companies behind them must be constrained in their abilities to process, use and disseminate individual-related information. The same limitations must be applied to private organizations and public authorities trying to access the information in social networks to use it for their own purposes.

\section{Theories of Privacy and Data Protection}

The history of privacy regulation can be traced back to ancient times. The Hippocratic oath is one of the oldest examples, protecting the confidentiality of all private information about the patient becoming known to her physician. Since at least the Fourth Council of the Lateran 1215 the clergy must protect the secrecy of confession. The bank secrecy is accepted since the seventeenth century, at least in Germany. All of these are based on a contractual or quasi-contractual protection of information entrusted to the care of specific professions. In addition the Roman law knew with the \emph{actio iniuriarum} a protection against specific indiscretions also outside of special trusted relationships \cite{Maass:1970}.

In the end of the nineteenth century the modern privacy debate started almost simultaneously in the US and in Europe. Based on Roman legal tradition, French and British precedents, and the history of the copyright Josef Kohler formulated the right of the author to decide whether to publish or not her sensations and feelings in writing as a special case of a general \emph{Individualrecht} (right of the individual) envisioned by him \cite{Kohler:1880}. The very same argumentation was used by Samuel D. Warren and Louis D. Brandeis in their seminal paper on the right of privacy \cite{WarrenBrandeis:1890}. Their \emph{right to be let alone} was just an update of Kohler's \emph{noli me tangere} in light of the development of the instantaneous photography and the emergence of the yellow press. Otto Gierke then generalized and reformulated Kohler's \emph{Individualrecht} as the \emph{allgemeines Persönlichkeitsrecht} (general right of personality) \cite{Gierke:1895}.

The right to privacy was quite successful used as a term and a concept in the American common law albeit there was no consensus of what it means and how it must be treated. After many extremely varying court decisions William L. Prosser argued that behind the right to privacy there are in fact four different interests against four different intrusions: (1) the intrusion upon the individual's seclusion or solitude, or into her private affairs, (2) public disclosure of embarrassing private facts about the individual, (3) publicity which places the individual in a false light in the public eye, and (4) appropriation, for the intruder's advantage, of the individual's name or likeness \cite{Prosser:1960}. Prosser was heavily criticized by Edward J. Bloustein for his distorting citations, improper classifications, and untenable conclusions \cite{Bloustein:1964}. While many scholars followed Bloustein in his holistic approach to regard privacy as an aspect of human dignity, the American legislator and most courts followed Prosser's more pragmatic approach. Prosser also heavily influenced the upcoming privacy debate in the computer science field and most of their technical approaches to implement privacy in electronic data processing systems.

\noindent
In Germany, the legislator and the jurisprudence for a long time did not recognize a general right of personality as a legal concept. Instead they protected specific expressions of this right as independent rights, like the \emph{Recht am eigenen Bild} (right to one's own picture), or the \emph{Recht am gesprochenen Wort} (right to the spoken word). After the \emph{Bonner Grundgesetz} (German Basic Law) came into effect the \emph{Bundesgerichtshof} (Federal Court of Justice) acknowledged the general right of personality as constitutionally protected (Article 2 (1) in conjunction with Article 1 (1) Basic Law). Since the middle of the 1950s this is settled case law of the \emph{Bundesverfassungsgericht} (Federal Constitutional Court), too. The general right of personality is therefore equally protected in constitutional law as in common law while its specific expressions might be protected in either the constitutional law, or the common law, or both. For more information about similarities and differences between the German right of personality and the American right of privacy at the end of the pre-digital era see \cite{Krause:1965}, \cite{Stroemholm:1967}, and \cite{Kamlah:1969}.

The public dispute about the National Data Center in the mid-60s can be considered as the starting point of the modern information privacy and data protection debate \cite{Dunn:1967}. Besides a strong focus on data quality and safety requirements the debate centered around guaranteeing secrecy and confidentiality for collected information, and some restrictions on how to collect information. Privacy was first and foremost used in its meaning of a private sphere, not as a right of personality \cite{Miller:1969}, \cite{Simitis:1971}. The individual's need for such a private sphere was often justified with a psychological rationale. The main distinction was between information being private or being public. Public information should not need any protection because it is not private. While this concept of privacy lost ground in the legal field after the beginning of the 1970s the computer science scholars still based their research on the distinction of the private and the public. For a few years the public discourse is based again on this outdated concept, most notably with respect to social networks.

The second main individualistic concept of information privacy and data protection takes the more holistic approach based on the general right of personality, and an equal understanding of the right to privacy. The predominant functions of information privacy according to this view are individual self-determination, personal autonomy, and limited and protected communication \cite{Westin:1967}. Privacy is not viewed as limited to secrecy \cite{Shils:1966}, \cite{Karst:1966}. Instead it means the control over information about oneself \cite{Westin:1967} or informational self-determination \cite{BVerfGE:65:1}. Most modern (European) data protection laws are based on this concept: The individual must know who knows what about her, and should retain some control over the collection, the processing, and the dissemination of information about her even while the information itself is owned by the data processor \cite{95:46:EC}, \cite{Tinnefeld:2005}.

While most scholars only examined the consequences of electronic data processing on the individual, others also considered the effects on structural aspects of society. Jeffrey A. Meldman noticed that Americans have always rather tolerated inefficiency than permitted the occurrence of unchecked power, particularly if centralized \cite{Meldman:1969}. Malcolm Warner and M. G. Stone also warned of possible bureaucratic omnipotence stemming from a broad employment of computers to process data \cite{StoneWarner:1969}. Even legislators paid attention to the broader problem. The first data protection law in the German state of Hesse contained rules to protect the balance of power between the legislative and the executive body of the state \cite{HDSG:1970}. Adalbert Podlech noticed first, as far as I can see, that for the formulation of a holistic data protection law a theory is needed encompassing the consequences both on the individual and the societal institutions \cite{Podlech:1970}. Concepts borrowed from both Niklas Luhmann's works \cite{Luhmann:1964a}, \cite{Luhmann:1986}, \cite{Luhmann:1966}, \cite{Luhmann:1967}, \cite{Luhmann:1969}, and \cite{Luhmann:1972} about his sociological systems theory and state organization law theories were used to describe the social function of data protection \cite{Podlech:1972}, \cite{Steinmueller:1972}, \cite{vBerg:1972}, \cite{Dammann:1974:Scheuch}. While in general Luhmann's systems theory lost the scientific discourse against more modern sociological theories, it nevertheless provided the basis for some important new directions in data protection like \emph{Systemdatenschutz} (system data protection) \cite{Podlech:1982}, data protection conformity in system design \cite{Systemgestaltung:1990}, or identity management \cite{Bizer:2004:Rost}.

\section{Functional Differentiation and the Social Function of Data Protection}

Most important is the concept of functional differentiation \cite{Luhmann:1967}, \cite{Luhmann:2004}, and its consequences on social systems and individuals alike. Functional differentiation means that a social system differentiates itself from an environment through the function it performs for the overall system \cite{Luhmann:1998}. That means different systems are distinguishable from each other by the different functions they perform. This separation of duties is often being used as a protection measurement, i.e. the separation of church and state, or the separation of powers in modern, democratic states to balance their powers and protect society. The functional differentiation thus parallels the division of labor, with its reasons as well as with its consequences. Losing one of them modern societies cannot be sustained.

On the societal level Luhmann distinguishes general social systems like Science, Politics, Economics, Law, or Religion. Each of these systems is using its own defining binary code: Science uses true vs.\ false, Politics uses power vs.\ no power, Economics uses payment vs.\ nonpayment, etc. That means for example that in Science only scientific truth matters, neither money nor power. If you can buy scientific truth with money or enforce it with power, you have no Science.

The correspondent to the functional differentiation on the state level in western democracies is the separation of power. This separation is applied horizontally as well as vertically. Horizontally, the \emph{trias} consists in the legislature, the executive, and the judiciary. They serve different functions: The legislature makes the law, the executive acts under the law, and the judiciary controls the executive actions with respect to the law. Vertically, modern states are differentiated in a municipal level, a state or intermediate level, and the national level. Using both these differentiations we are able to balance the powers even as states as a hole are much more powerful entities than in the past \cite{Dammann:1974:SchimmelSteinmueller}.

The executive body itself became historically more and more differentiated, too. The many authorities perform different duties using different means. One example is the separation of the police and the social welfare administration, previously being part of the \emph{Polizey}. Much later, the same happened with the separation of the police and the registry office, or the separation of the police and the intelligence service \cite{Lewinski:2009}. In a constitutional state the police acts primarily repressive while the intelligence service may also act preventative, nevertheless it is not allowed to arrest people. There is no \emph{Einheit der Verwaltung} (unity of authority) as in an absolutistic state---the different authorities are structurally and informationally separated, and therefore limited in their power.

Before the emergence of digital computers, data processing and information processing was done by humans, and therefore slow, inefficient and error-prone. The computer tends to revoke all limitations of manual data processing, eases the processing of information, and therefore undermines the protective character of inefficiency \cite{Kilian:1973:Steinmueller}. Because information serve the production of decisions whoever controls the collection and processing of information controls the decisions based thereon. External entities like the parliament, the data protection commissioner, or even more the data subject are structurally unable to control the data processor. The control of the processing of information becomes more and more centralized even if the computing itself becomes decentralized or distributed. Local authorities and local democratic institutions therefore tend to lose power to centralized ones.

The functional differentiation not only has consequences for social systems or the society but also for the individual. In modern differentiated societies the individual plays different roles in different contexts. Although there exist expectations on how to behave in specific contexts the individual autonomously decides how to play her roles as sister, friend, neighbor, colleague, principal, patient, business partner, client, voter, tax payer, etc. Every human being and every social system she interacts with only sees the individual in her current role. The roles are generally separated, the individual---the totality of her roles---only being known to the individual itself. Her role-playing is basis for and product of her right to personality \cite{Dammann:1974:Mueller}. Social players being able to consolidate information from different roles based on the abilities of modern data processing technology, ubiquitous data collection, and sophisticated data mining methods would threaten the individual with total visibility and transparency. The data processor would be able to predict the future role-playing of the individual and to base its decisions on this information advantage. This information power threatens the autonomy of the individual \cite{Mueller:1975}.

The aim of data protection is therefore (1) to maintain the functional differentiation of society against the threats posed by modern information processing, and (2) to counter undue information power by organized social players. Data protection guarantees the balance of power between different social systems, societal institutions, and other social groups, and protects the role-playing of the individual and therefore her autonomy. This is done by controlling the flow of information between individuals, institutions, and different sectors of society. Therefore data protection is the controlled assignment or retention of information to prevent socially undesirable information processing and to limit organizational power over individuals, groups, social systems, and society.

\section{Social Networks and Data Protection}

Social networks like Facebook or Google+ often get criticized for exposing people to the public gaze against their will, or for helping criminals spying or stalking by making ``private'' data publicly available. So called privacy critics counter with pointing on the consent of the individuals, or by claiming that information given to the social networks are public, and therefore not deserving protection. But as shown before these are not the problems data protection tries to prevent. They are either IT security problems, or based on outdated privacy theories.

Instead, the main problem of social networks is their ability to collect information on different roles of the individual and merge them into one holistic and exhaustive picture. Because this modeling of the individual is not only based on information provided by the individual itself but also based on information provided by other people or intrinsic properties of the social graph around each one, the ability of the individual to control what parts of her personality is known to the owner of the social networking service is seriously limited. Most people are not even aware of how much information may be deduced from the social graph, see for example \cite{Jernigan:2009}. The roles being made transparent cover all areas of life: family, education, work life, hobbies, and even politics. With their members made transparent, informal groups of people, formal associations, companies, and even institutions become transparent, too. In addition more and more groups, associations, and institutions use social networks for their own internal or external communications.

Second, there are usually no limitations for the company behind the social network to process and use all information, either collected or derived. There are also almost no limitations on handing over information from or about the individual to government authorities, or private organizations. For example, Facebook's privacy policy reads: ``We may also share information when we have a good faith belief it is necessary to prevent fraud or other illegal activity, to prevent imminent bodily harm, or to protect ourselves and you from people violating our Statement of Rights and Responsibilities. This may include sharing information with other companies, lawyers, courts or other government entities,'' (cited after \cite{Semitsu:2011}). What the social network knows, the state knows, too.

With individuals, groups, and social systems made visible and transparent alike, and widespread sharing of information between social networks and primarily government authorities, the balance of power is shifted in favor of centralized bureaucracies, either private or public. Legislation loses its ability to control the executive if the latter acts \emph{in arcanum}, and the former is being made transparent and therefore predictable. Individuals, groups, and associations alike lose their autonomy---and with it their freedom---against businesses and public authorities. The functional differentiation as a limitation of power of social systems in modern western societies may collapse---at least in some areas---if there exist social players being able to transgress informational boundaries between social systems. The modern state as the most powerful member of society may become total again.

From a data protection point of view there would be two fundamental claims: (1) General social networking services must be information sinks concerning information about individuals and groups. They should not be allowed to disseminate individual-related information to other organizations, either private or public. In the same way as the inviolability of the home (Article 13 (1) Basic Law) or the \emph{Grundrecht auf Gewährleistung der Vertraulichkeit und Integrität informationstechnischer Systeme} (right to the provision of confidentiality and integrity of information technology systems) \cite{BVerfGE:120:274} protect the individual and her personality through the protection of structures (the home and the computer, respectively), information stored in social networks must be protected as a whole due to their ability to provide a total picture of the individual. There also must be a general prohibition of retrieving and using such information by public authorities, especially police and intelligence services. (2) For structurally limiting the power of companies behind social networks, they should be treated and regulated like monopolies. First of all, it must be prevented that they use their informational power over their users to gain a hold in other areas, especially politics or in collaboration with government agencies.

\section{Conclusion}

General social networks should not only be criticized for exposing their users to the public, for their low privacy and security standards, and their incomprehensible privacy policies. Based on sociological theories of Niklas Luhmann and the pioneering works of Adalbert Podlech, Paul J. Müller, and others concerning the foundations of data protection I have shown in this paper that general social networks and their ability to collect, store, and process vast amounts of information about individuals, groups, and organizations from different social systems and to combine them in centralized data banks constitute a threat (1) to the individual's ability to control her own role-playing---and therefore her autonomy and freedom---in modern, functionally differentiated societies, (2) to groups, associations, and social systems being dependent on functional differentiation as protection against overly powerful public or private entities, and (3) to the functionally differentiated society as a whole with its dependency on a balance of power to guarantee freedom, democracy, and a state of law.

\paragraph{Acknowledgements.}

The author would like to thank Martin Rost for providing the idea to study the consequences of social networking services on data protection from a structuralistic point of view. He also wishes to thank Martin Warnke, Wolfgang Coy, and especially Jochen Koubek for enlightening and fruitful discussions.

\printbibliography

\end{document}